\documentclass{appolb}
\usepackage{graphicx}

\begin{document}
\title{Design of the SABAT system for underwater detection of dangerous substances%
\thanks{Presented at the Jagiellonian Symposium of Fundamental and Applied Subatomic
Physics, Krak{\'o}w, 11.06.2015}%
}
\author{M.~Silarski$^{a}$, D.~Hunik$^{a}$, M.~Smolis$^{a}$, S.~Tadeja$^{a}$, P.~Moskal$^{a}$
\address{$^{a}$Institute of Physics, Jagiellonian University \\ {\L}ojasiewicza 11, 30-348 Krak\'ow, Poland}
}
\maketitle
\begin{abstract}
We present status of  simulations used to design a novel device for the detection
of hazardous substances in the aquatic environment using neutron activation.
Unlike the other considered methods based on this technique we propose to use guides
for neutron and gamma quanta which speeds~up and simplifies identification. First
preliminary results show that both the neutron guide and the $\gamma$ ray guide
increase the performance of underwater threats detection.
\end{abstract}
\PACS{P82.80.Jp, 89.20.Dd}
  
\section{Introduction}
Growing risk of terrorism demands a constant development of new techniques for
hazardous substances detection. One of the best methods for homeland security are based
on the Neutron Activation Analysis (NAA) providing non-intrusive detection of explosives or drugs.
The principle of the method is to use a fast neutron beam produced using sealed D+T generator
to excite nuclei of investigated substance and detect characteristic gamma quanta produced
in de-excitation of the nuclei. This allows one to identify the stoichiometry of the substance
and determine if it is dangereus~\cite{moskalAnn,sabat1}. So far there are several devices based
on Neutron Activation Analysis designed for ground homeland security which were introduced
in the USA by Science Applications International Corporation~\cite{phoenix} and CALSEC~\cite{maglich},
and in Europe in Sodern~\cite{sodern}, EURITRACK~\cite{euritr} and SWAN~\cite{swan} projects.
These devices are used in the homeland security or contraband detection on the land.
However, in an aquatic environment one encounters serious problems due to strong attenuation
of neutrons in water. Moreover, as in the case of ground detectors, an isotropic generation
of neutrons induces a large environmental background. This noise can be significantly reduced
by the requirement of the coincident detection of the alpha particles which are produced together
with neutrons~\cite{viesti1,viesti_phd}. The attenuation of neutrons can
be compensated by reducing the distance between generator and examined item~\cite{uncoss2}.
There are also solutions based on low energy neutrons which are moderated in
water before reaching the tested object. The detector is then counting the gamma quanta from
thermal neutron capture and secondary neutrons originating from the irradiated object.
The identification is done by searching for anomalies in the observed spectra of gamma quanta
and neutrons~\cite{ActaA}. However, these methods do not allow to detect explosives buried
deeper in the bottom of the sea and strong attenuation of neutrons and gamma quanta
significantly increases the exposure time.\\
An alternative solution of a detector which uses NAA technique and special guides for
neutrons and emitted gamma rays was proposed within the SABAT project~\cite{patent1}.
The device allows for detection of dangerous substances hidden deep in the bottom of the sea
with significantly reduced background and provides determination of the density distribution
of the dangerous substance in the tested object~\cite{ActaA}. In this article we present
status of the design of the SABAT detector based on Monte Carlo simulations, focusing on the
impact of the usage of neutron and gamma quanta guides.
\section{Status of the simulation}
In order to optimize the dimensions and relative positions of detectors and guides we have
developed dedicated open source software package written in the C++ programming language~\cite{ActaA}.
This simulation tool is using novel methods of geometry definition and particle tracking
and the Open MPI library supporting parallel computing~\cite{mpi}.
To define the geometry of simulated objects we use a polygon mesh, which is in the simplest form
a collection of vertices and edges that are used to define a surface of the polyhedral 3D object.
In our case the faces consist of triangles. Trajectory of a particle traveling in a scene
consisting of such defined objects is simulated with so-called ray tracing technique. To determine
if a particle entered given object the M\"oller-Trumbore method was implemented which searches for
intersection of the particle trajectory with the object surface (triangle mesh).
The total cross-sections and angular distributions of neutrons and $\gamma$ rays interaction
with selected nuclei were parametrized as a function of energy using data from the Evaluated Nuclear
Data File (ENDF)
database~\cite{endf}, while gamma quanta energies were taken from the Evaluated Nuclear Structure
Data Files (ENSDF)~\cite{ensdf}.
\begin{figure}
\centering
\includegraphics[width=5.50cm]{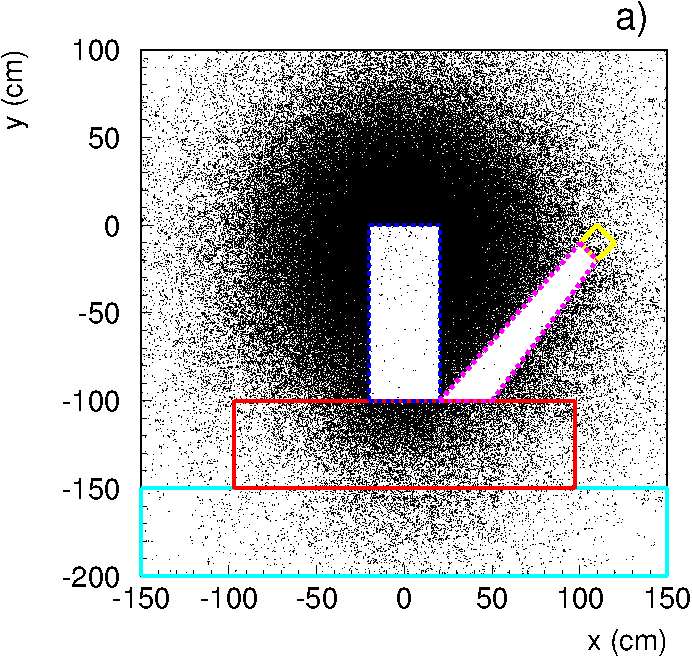}
\includegraphics[width=5.50cm]{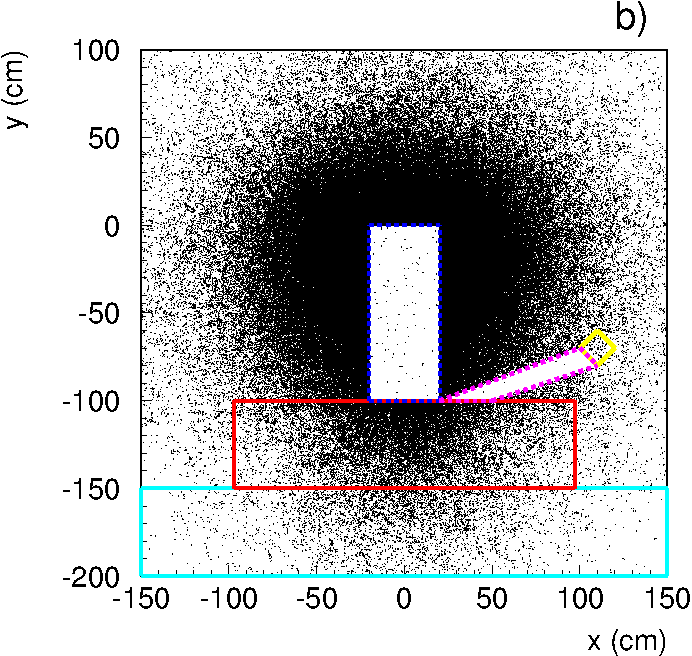}
\includegraphics[width=5.50cm]{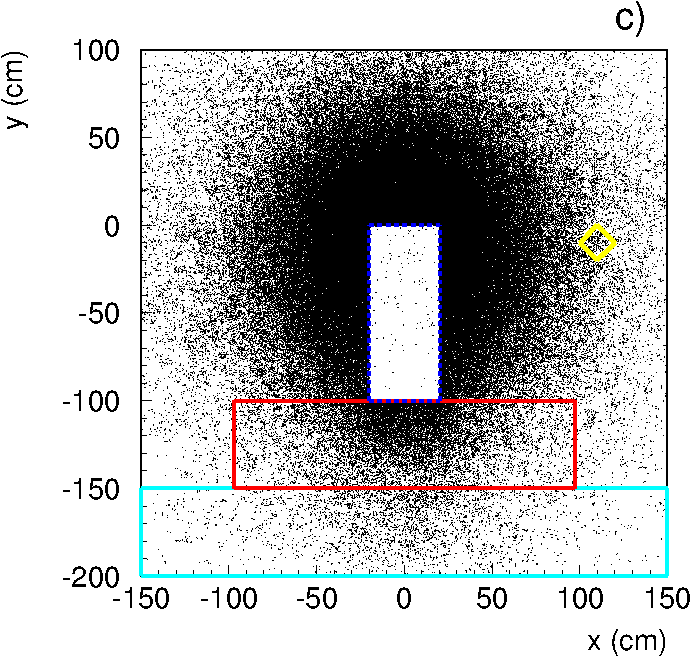}
\includegraphics[width=5.50cm]{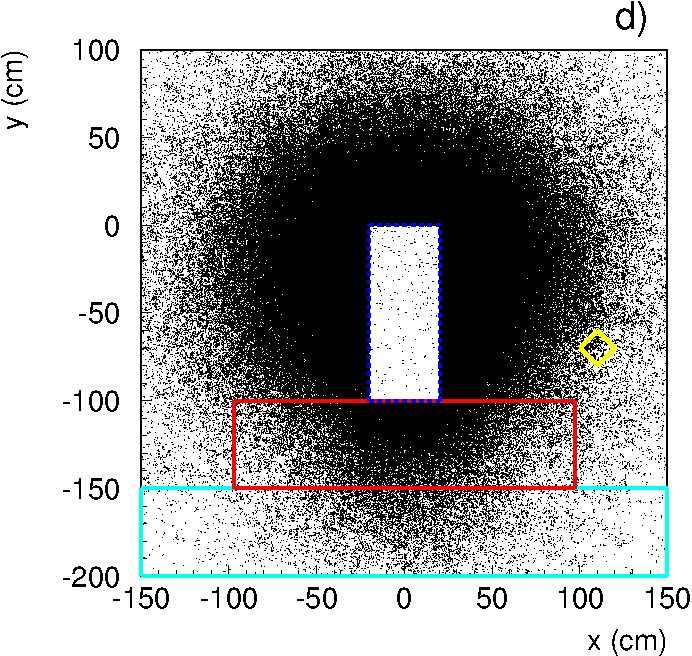}
\caption{Distribution of $\gamma$ quanta interaction points in x-y views
for two positions of the detector with (a and b) and without gamma quanta guide (c and d).
The point-like source of neutrons is located at the origin of the reference frame.
The neutron guide is presented as the blue dashed rectangle aligned along the y axis.
The container with mustard gas is a red cuboid aligned horizontally which lies at the bottom
of the sea marked in cyan. The Ge detector (in yellow) is connected to the gamma rays guide
(dashed magenta polyhedron). For better visibility of the influence of guides plots are made
only for -10~$\leq$~z~$\leq$~10 cm.
}
\label{Fig1}
\end{figure}
As a starting point for design of the device for underwater threats detection we have defined
a simple setup with point-like source generating uniformly in space 14.1~MeV neutrons in the 
neutron-$\alpha$ center of mass frame. As a first step, to study the influence on the measurement
of gamma
quanta guide and the relative position of generator and detector we have simulated 1.4$\cdot$10$^7$
events with two different detector locations with and without gamma rays guide. The scheme
of the simulated setups with superimposed $\gamma$ quanta interaction points are shown in
Fig.~\ref{Fig1}.
The interrogated object with dimensions 194~x~50~x~50~cm$^3$ lies at the bottom of a sea and
contains mustard gas (C$_4$H$_8$Cl$_2$S). The neutron guide is a cuboid with dimensions
40~x~100~x~40~cm$^3$, while gamma rays guides is represented by polyhedron with 15~x~20 cm$^2$
and 30~x~30~cm$^2$ bases.
They are both filled with air under normal conditions. As a detector we have simulated
15~x~15~x~20~cm$^3$ Ge crystal measuring the energy of gamma rays with  6\% resolution (FWHM).
The detector surface with dimensions of 15 x 20 cm$^2$ is adjusted to the shape of the gamma
quanta guide.
\begin{figure}
\centering
\includegraphics[width=5.50cm]{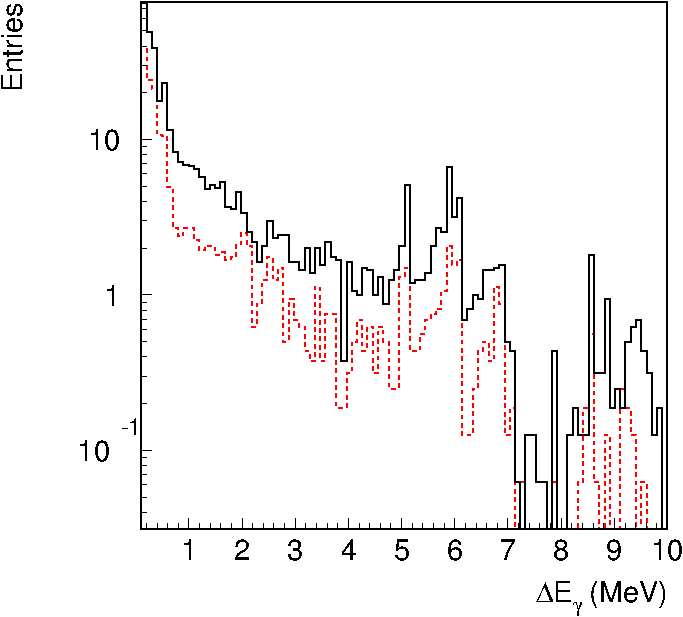}
\caption{Simulated energy spectra as would be measured with the germanium detector. Simulations
were performed for the setups presented in Fig.~\ref{Fig1}b
 (solid histogram) and in Fig.~\ref{Fig1}d (dashed histogram). The small signal from the mustard
gas can be seen at about 9.6~MeV and 4.4~MeV (carbon), 2.2~MeV (sulfur) and around 6.5~MeV (chlorine).
The huge background  is mainly due to the gamma quanta from neutron-oxygen reactions.
The energy  of these quanta is smeared due to the scattering in the water.
The number of entries is normalized to the generated statistics.
}
\label{Fig2}
\end{figure}
The exemplary energy spectra measured
by the germanium detector for one detector position are shown in Fig.~\ref{Fig2}. As one can see
in Tab.~\ref{Tab1}, among all of the gamma quanta reaching the detector about 80$\%$ originate
from oxygen in the water independently of the detector position or presence of the guide.  
\begin{table}
\caption{Fractional composition of sources of gamma quanta registered by the detector
(in $\%$) for all the simulated cases shown in Fig.~\ref{Fig1}. The last column shows
the overall fraction of gamma quanta reaching the detector and originating from
the mustard gas.}
\label{Tab1}
\begin{tabular}{|c|c|c|c|c|c|c|c|c|c|c|}
 \hline
Setup & H & C & N & O & Na & Si & S & Cl & n & \begin{tabular}[c]{@{}c@{}}Mustard\\Gas\end{tabular}\\
\hline
a     & 0 & 0.77 & 0 & 85.17 & 0.24 & 0 & 0 & 0.26 & 13.56 & 1.02\\
\hline
b     & 0.53 & 1.68 & 0.10 & 76.50 & 0.38 & 0.08 & 0.63 & 1.78 & 18.32 &3.30\\
\hline
c     & 0.59 & 0 & 0 & 81.71 & 0.88 & 0 & 0 & 0.89 & 15.93 &0\\
\hline
d     & 7.88 & 0.59 & 0.05 & 77.04 & 0.69 & 0.24 & 0.10 & 6.95 & 6.46 &2.25\\
\hline
\end{tabular}
\end{table}
The second relevant source of background constitute neutrons which scatter in water and reach
the detector. The amount of registered gamma quanta which were created in the
container with mustard gas is quite small but the advantage of using guide is visible.
The mustard gas is composed of carbon, hydrogen, chlorine and sulfur.
However, Fig.~\ref{Fig2} indicates that the 4.4~MeV carbon line is overwhelmed by background while
we see clearly gamma quanta from carbon with energy of about 9.6~MeV. In case of chlorine and
sulfur (lines of 6.5~MeV and 2.2~MeV, respectively) the oxygen background has to be
significantly reduced to allow one to identify these elements more clearly. This can be done
requiring of the coincident detection of the $\alpha$ particle generated together with neutron
or by covering all
the detector faces not connected to the gamma quanta guide with thick layer of absorber.
Identification of sulfur and chlorine is difficult due to neutron capture on hydrogen giving
lines of very close energy and chlorine content in the sea water~\cite{ActaA}.
However, using the gamma quanta guide one can detect $\gamma$ rays from these two elements not
affected by scattering in the water which gives a great advantage in the identification.  
\section{Conclusions and outlook}
In the framework of the SABAT project we have been developing simulation package based on
novel methods of geometry definition and particle tracking. Although we are still in a very
early stage of development the first results indicate that indeed, both the neutron guide and
$\gamma$ ray guide will increase the performance of underwater threats detection with fast
neutrons~\cite{ActaA}. With 1.4$\cdot$10$^7$ generated neutrons we see enhancements of signal
events in the detector
together with huge background originating mostly from oxygen, as it was expected. This shows clearly
that all the faces of the detector not connected to the  guide should be covered with thick
layer of material absorbing $\gamma$ quanta and neutrons which constitute another big source of
background. This noise can be also significantly reduced by the requirement of the coincident
detection of the $\alpha$ particle generated together with neutron, which allows for the neutron
tagging~\cite{ActaA}. As one of the possible way of the background suppression via time-of-flight
method we consider an application of plastic scintillators in order to take advantage of their excellent
timing properties, as recently demonstrated in case of low energy gamma
quanta~\cite{jpet2014_1,jpet2014_2,jpet2015_1,jpet2015_2}.
The $\alpha$ particle detection and time measurement will be included as the next
step in the development of the simulations. This will allow for a final design of the device
for underwater threats detection with fast neutrons, which requires much bigger statistics
generated.

\end{document}